\documentclass[english]{article}
\usepackage[T1]{fontenc}
\usepackage[latin9]{inputenc}
\usepackage{geometry}
\usepackage{color}

\makeatletter
\@ifundefined{definecolor}
 {\usepackage{color}}{}
\@ifundefined{definecolor}
 {\@ifundefined{definecolor}
 {\usepackage{color}}{}
}{}
\@ifundefined{definecolor}
 {\@ifundefined{definecolor}
 {\@ifundefined{definecolor}
 {\usepackage{color}}{}
}{}
}{}
\makeatother

\makeatother

\usepackage{babel}

\makeatother

\usepackage{babel}

\makeatother

\usepackage{babel}

\makeatother

\usepackage{babel}

\begin{document}

\title{Beyond the Goldenberg-Vaidman protocol: Secure and efficient quantum
communication using arbitrary, orthogonal, multi-particle quantum
states}

\author{Chitra Shukla$^{1}$, Anirban Pathak$^{1,2}$ and R. Srikanth$^{3,4}$}

\maketitle
\begin{center}
$^{1}$Jaypee Institute of Information Technology, A-10, Sector-62,
Noida, India. 
\par\end{center}

\begin{center}
$^{2}$RCPTM, Joint Laboratory of Optics of Palacky University and 
\par\end{center}

\begin{center}
Institute of Physics of Academy of Science of the Czech Republic,
Faculty of Science, Palacky University, 17. listopadu 12, 77207 Olomouc,
Czech Republic. 
\par\end{center}

\begin{center}
$^{3}$Poornaprajna Institute of Scientific Research, Sadashivnagar,
Bengaluru- 560080, India. 
\par\end{center}

\begin{center}
$^{4}$Raman Research Institute, Sadashivnagar, Bengaluru- 560060,
India. 
\par\end{center}
\begin{abstract}
It is shown that maximally efficient protocols for secure direct quantum
communications can be constructed using any arbitrary orthogonal basis.
This establishes that no set of quantum states (e.g. GHZ states, $W$
states, Brown states or Cluster states) has an advantage over the
others, barring the relative difficulty in physical implementation.
The work provides a wide choice of states for experimental realization
of direct secure quantum communication protocols. We have also shown
that this protocol can be generalized to a completely orthogonal state
based protocol of Goldenberg-Vaidman (GV) type. The security of these
protocols essentially arises from duality and monogamy of entanglement.
This stands in contrast to protocols that employ non-orthogonal states,
like Bennett-Brassard 1984 (BB84), where the security essentially
comes from non-commutativity in the observable algebra. 
\end{abstract}

\section{Introduction}

Since the proposal for the first protocol for quantum key distribution
\cite{bb84}, and the other early QKD protocols \cite{bb84,ekert,b92,vaidman-goldenberg},
it came to be understood that quantum states can be employed for other
cryptographic tasks, for example, for quantum secret sharing (QSS)
of classical secrets \cite{Hillery}. A protocol for \textit{deterministic
secure quantum communication} (DSQC) using entangled photon pairs,
was proposed by Shimizu and Imoto \cite{Imoto}. Although it was found
to be insecure, it suggested that the prior generation of key (i.e.
QKD) can be circumvented and protocols for unconditionally secure
direct quantum communication of the message can be designed. Subsequently,
many such protocols were proposed, which can broadly be divided into
two classes: (a) those for quantum secure direct communication (QSDC)
\cite{Long   and   Liu,ping-pong,lm05} and (b) those for DSQC \cite{dsqc_summation,dsqcqithout-max-entanglement,dsqcwithteleporta,entanglement      swapping,Hwang-Hwang-Tsai,reordering1,the:cao and song,the:high-capacity-wstate}.

In DSQC receiver (Bob) can read out the secret message only after
receipt of a \textit{pre-key}: additional classical information of
at least one bit for each qubit transmitted by the sender (Alice).
By contrast, when no pre-key is required, a secure communication protocol
is referred to as QSDC protocol \cite{review}. A conventional QKD
protocol generates the unconditionally secure key by quantum means
but then uses classical cryptographic resources to encode the message.
No such classical means are required in DSQC and QSDC. This interesting
feature of DSQC and QSDC protocols has motivated several groups to
study different aspects of DSQC and QSDC protocols in detail {[}\cite{review}
and reference therein{]}.

QKD can be obtained from DSQC and QSDC protocols in the sense that
given local resources like a random number generator, Alice can always
implement QKD if she can implement DSQC or QSDC. Consequently, if
we can establish that DSQC and/or QSDC protocol can be realized by
using an arbitrary set of linearly independent quantum states, that
would imply that QKD can also be realized using an arbitrary set of
linearly independent quantum states. This is a motivation to study
this type of protocols in the present paper.

The unconditional security of the existing protocols is claimed to
be obtained by using different quantum resources. For example, unconditionally
secure protocols are proposed a) with and without maximally entangled
state {[}\cite{dsqcqithout-max-entanglement} and references therein{]},
b) using teleportation \cite{dsqcwithteleporta} c) using entanglement
swapping \cite{entanglement  swapping} d) using rearrangement of
order of particles \cite{reordering1,the:C.-W.-Tsai} etc. Although
based on very different resources, the security of all the existing
protocols of secure direct quantum communication arises ultimately
from the use of conjugate coding (i.e. from quantum non-commutativity).
This is so because all the existing protocols of DSQC and QSDC detects
Eve by measuring verification qubits in 2 or more mutually unbiased
bases (MUBs). We thus classify all these protocols as essentially
of BB84 type.

Now one may ask: Is conjugate coding essential for DSQC and QSDC?
The answer is {}``no'', as we have recently shown the same in a
different context \cite{With preeti}. Here we will first propose
DSQC and QSDC protocols of BB84 type (i.e. one which uses conjugate
coding) using arbitrary quantum states and will subsequently turn
them to Goldenberg-Vaidman (GV) type \cite{vaidman-goldenberg} protocols,
which uses only orthogonal states for encoding, decoding and error
checking, as was done in the original GV protocol of QKD. So far the
GV protocol has existed as an isolated orthogonal-state-based protocol
of QKD. Our Bell-state-based generalization of original GV protocol
may also be regarded as DSQC or QSDC, thus providing, in our opinion,
the first instance of GV-type DSQC and QSDC protocols \cite{With   preeti}.
The idea is further extended here and it is shown that GV type protocols
of DSQC and QSDC can be implemented using an arbitrary set of linearly
independent quantum states.

In the protocols of DSQC and QSDC to be proposed in this paper, the
rearrangement of orders of particles, plays a very crucial role. Therefore,
it would be apt to note that the DSQC protocol based on the rearrangement
of orders of particles was first proposed by Zhu \emph{et al}. \cite{reordering1}
in 2006. However, it turns to be insecure under a Trojan-horse attack,
which can be corrected using the idea of rearrangement of particle
order, as noted by Li \emph{et al}. \cite{dsqcqithout-max-entanglement}.
The Li \emph{et al.} protocol may thus be considered as the first
unconditionally secure protocol of DSQC based on \textit{permutation
of particles} (PoP) .

In the last five years, many such PoP-based protocols of DSQC have
been proposed. Very recently, Banerjee and Pathak \cite{With   Anindita-pla},
Shukla, Banerjee and Pathak \cite{With chitra-ijtp}, Yuan \emph{et
al}. \cite{the:high-capacity-wstate} and Tsai \emph{et al}. \cite{the:C.-W.-Tsai}
have proposed PoP-based protocols for both DSQC and QSDC. The Yuan
\emph{et al}. protocol and Shukla-Banerjee-Pathak protocol use 4-qubit
symmetric $W$ state for communication, while the Banerjee-Pathak
protocol uses 3-qubit $GHZ$-like states, and the Tsai \emph{et al.}
protocol utilizes the dense coding of four-qubit cluster states.

Here we would also like to mention that in 2009 Xiu \emph{et al.}
\cite{XIU} had provided a protocol of DSQC using the 5-qubit Brown
state. Almost at the same time a DSQC protocol using the 5-qubit Brown
state was also proposed by Jain, Muralidharan and Panigrahi \cite{Sakshi}.
Thus we observe that secure direct quantum communication, which was
initially proposed using 2-qubit Bell states \cite{Long  and   Liu,ping-pong}
has subsequently been proposed with much more complex entangled states
like 3-qubit $GHZ$-like state \cite{With   Anindita-pla}, 4-qubit
$W$ state \cite{the:high-capacity-wstate,With   chitra-ijtp}, 4-qubit
cluster state \cite{the:C.-W.-Tsai}, and the 5-qubit Brown state
\cite{XIU,Sakshi}. The widespread use of these maximally entangled
states in DSQC, QSDC and QKD have by now established a common perception
that these states are special and are important for the implementation
of quantum cryptographic protocol.

Contrary to this common belief we will establish that DSQC and QSDC
is possible with any arbitrary quantum state. Further, earlier proposed
protocols are compared with each other and on the basis of qubit efficiency
one is claimed to be better over the others. Here we aim to show that
no quantum state is special and one can construct maximally efficient
DSQC and QSDC protocol starting from any arbitrary set of linearly
independent state vectors.

The remaining part of the present paper is organized as follows, in
Section \ref{sec:The-protocols-ofdsqc} we propose BB84-type protocols
for DSQC and QSDC using arbitrary quantum state. In Section \ref{sec:Generalisation-to-orthogonal},
we modify these protocols to corresponding orthogonal-state-based
protocols of GV type. In Section \ref{sec:Efficiency-and-security}
we analyze the security and efficiency of the proposed protocol and
Section \ref{sec:Conclusions} is dedicated for the conclusions.

\section{The protocols of DSQC and QSDC\label{sec:The-protocols-ofdsqc}}

Our aim is to show that efficient DSQC and QSDC protocols can be constructed
using an arbitrary set of linearly independent quantum states. However,
the information encoded states sent by Alice must be mutually orthogonal
in order to enable Bob to unambiguously distinguish the encoded information.
Let us assume that Alice starts with a set of $M\equiv2^{n}$ arbitrary,
linearly independent $n$-qubit state vectors $|x_{1}\rangle,\,|x_{2}\rangle,\,|x_{3}\rangle,\cdots,|x_{M}\rangle$.

If they are not pairwise orthogonal, Alice can apply Gram-Schmidt
procedure on the above set of state vectors and obtain a set of mutually
orthonormal state vectors as $\{|a_{1}\rangle,|a_{2}\rangle,|a_{3}\rangle,\cdots,|a_{M}\rangle\}$.
This forms our arbitrary basis set. Let $\Pi_{J}$ ($J=1,2,\cdots,M!$)
be an arbitrary permutation on $M$ letters, and let $\{|b_{j}\rangle\equiv|a_{\Pi(j)}\rangle\}$
represent the new basis obtained simply by permuting the states in
the first basis. Effectively, we are relabeling the states $|a_{j}\rangle$.
Now we may introduce the operators $U_{J}=\sum_{j}|b_{j}\rangle\langle a_{j}|$,
which are unitary, as is easily verified to satisfy $U_{J}U_{J}^{\dagger}=U_{J}^{\dagger}U_{J}=\left(\sum_{p}|a_{p}\rangle\langle b_{p}|\right)\left(\sum_{q}|b_{q}\rangle\langle a_{q}|\right)=\left(\sum_{j}|a_{j}\rangle\langle a_{j}|\right)=\mathcal{I}_{M}$.
In principle, if $\left\{ |b_{j}\rangle\right\} $ is any basis set
in $M$ dimension then $U_{J}$ will be unitary. However, for our
purpose of showing that DSQC/QSDC is possible with an arbitrary set
of mutually orthonormal state vectors $\{|a_{i}\rangle\}$ it is sufficient
to consider $\{|b_{j}\rangle\equiv|a_{\Pi(j)}\rangle\}$. Further,
this choice of output basis set is consistent with the spirit of the
original GV protocol, where only single basis set is used (MUBs are
not used) for encoding and decoding.

Thus $U_{J}$ are physically acceptable operators that can transform
one state vector of our arbitrary basis set $\{|a_{j}\rangle\}$ into
a state of the same basis set as basis set $\{|b_{j}\rangle\equiv|a_{\Pi(j)}\rangle\}$.
This is physically expected as these operators are nothing but rotation
in the state space spanned by the basis set $\{|a_{j}\rangle\}$.
Now we may define an orthogonal family of $M$ unitaries $\left\{ U_{j}\right\} \subset\left\{ U_{J}\right\} $
as one that satisfies $\langle a_{i}|U_{j}^{\dagger}U_{k}|a_{i}\rangle=0$,
where $|a_{i}\rangle\in\{|a_{j}\rangle\}$ would be used later as
the initial state. Given a particular permutation of vectors $|b_{j}\rangle\equiv|a_{\Pi(j)}\rangle$,
we construct our family of unitary operators $\left\{ U_{j}\right\} $
as any set of $M$ operators that yield $|b_{j}\rangle=U_{j}|a_{i}\rangle$
and satisfy $\langle a_{i}|U_{j}^{\dagger}U_{k}|a_{i}\rangle=0$. 

There are various alternative ways to construct $\left\{ U_{j}\right\} .$
For example, one particularly symmetric (in fact Hermitian) set $\left\{ U_{j}\right\} $
for the case $\{|b_{j}\rangle\equiv|a_{j}\rangle\}$ is provided below
as a specific example \begin{equation}
\begin{array}{lcl}
U_{i} & = & I,\\
U_{j\neq i} & = & |a_{i}\rangle\langle a_{j}|+|a_{j}\rangle\langle a_{i}|+\sum_{k=1;k\neq i,j}^{M}|a_{k}\rangle\langle a_{k}|.\end{array}\label{eq:unitary1}\end{equation}
 Such operator families are used by Alice for encoding classical information.

It is worth observing here that what was noted above and the protocols
below do not depend on whether the basis states $|a_{j}\rangle$ are
single-particle or multi-particle states. The only requirement is
that it is possible to split every state $|a_{j}\rangle$ into $n$
geometrically separable pieces that carry no information of the full
state. If they are single-particle states, then $|a_{j}\rangle$ is
an equal-weight superposition that lives in an $n$ (or greater) dimensional
space, and constructed for $n$ spatially separated wave-packets. 

Alice publicly announces the set of linearly independent states $\{|a_{j}\rangle\}$
to be used by her, the initial state $|a_{i}\rangle$ she is going
to prepare and the set of unitary operators $\left\{ U_{j}\right\} $
to be used by her for encoding. For the practical implementation purpose,
our assumption is that Alice has devices to prepare states in the
basis $\{|a_{j}\rangle\}$, to implement the set of unitary operators
$\left\{ U_{j}\right\} $, and that Bob has devices to make measurements
in $\{|b_{j}\rangle\}$ basis. The main part of the protocol of DSQC
works as follows.

\subsection*{The DSQC protocol}
\begin{description}
\item [{DSQC1:}] Alice prepares $|a_{i}\rangle^{\otimes N},$ which is
an $Nn$ qubit state (as $|a_{i}\rangle$ is $n$ qubit state). Qubits
of $|a_{i}\rangle^{\otimes N}$ are indexed as $p_{1},p_{2},\cdots,p_{Nn}.$
Thus $p_{s}$ is the $s^{{\rm th}}$ qubit of $|a_{i}\rangle^{\otimes N}$
and $\{p_{nl-n+1},p_{nl-n+2},\cdots,p_{nl}:l\leq N\}$ are the $n$
qubits of the $l^{{\rm th}}$ copy of $|a_{i}\rangle$. 
\item [{DSQC2:}] Alice encodes her $n$-bit classical secret message by
applying one of the $n$-qubit unitaries $\left\{ U_{j}\right\} =\left\{ U_{1},U_{2},\cdots,U_{M}\right\} .$
The encoding scheme, which is predefined and known to Bob, is such
that $U_{1},\, U_{2},\, U_{3},\cdots,U_{M}$ are used to encode $0_{1}0_{2}\cdots0_{n},\,0_{1}0_{2}\cdots1_{n},\,0_{1}0_{2}\cdots1_{n-1}0_{n},\cdots,1_{1}1_{2}\cdots1_{n}$
respectively. Thus the coded states $U_{j}|a_{i}\rangle=|b_{j}\rangle$
are mutually orthogonal. 
\item [{DSQC3:}] Using all the qubits of her possession, Alice creates
an ordered sequence $P_{B}=\left[p_{1},p_{2},p_{3},p_{4},\cdots,p_{Nn-1},p_{Nn}\right]$.
She prepares $Nn$ decoy qubits%
\footnote{When $2x$ qubits (a random mix of message qubits and decoy qubits)
travel through a channel accessible to Eve and $x$ of them are tested
for eavesdropping then for any $\delta>0,$ the probability of obtaining
less than $\delta n$ errors on the check qubits (decoy qubits), and
more than $(\delta+\epsilon)n$ errors on the remaining $x$ qubits
is asymptotically less than $\exp[-O(\epsilon^{2}x)]$ for large $x$.
\cite{Nielsen     chuang-589}. As the unconditional security obtained
in quantum cryptographic protocol relies on the fact that any attempt
of Eavesdropping can be identified. Thus to obtain an unconditional
security we always need to check half of the travel qubits for eavesdropping.
This is why $Nn$ decoy qubits are prepared. Decoy qubits can be prepared
in any two MUBs.%
}\textcolor{red}{{} }$d_{i}$ with $i=1,2,\cdots,n$ such that $d_{i}\in\{|0\rangle,|1\rangle,|+\rangle,|-\rangle\}$\textcolor{red}{{}
}and concatenates them with $P_{B}$ to yield a larger sequence $P_{B^{\prime}}=\left[p_{1},p_{2},p_{3},p_{4},...,p_{Nn-1},p_{Nn},d_{1},d_{2},d_{3},d_{4},...,d_{Nn-1},d_{Nn}\right]$.
Thereafter Alice applies a permutation operator $\Pi_{2Nn}$ on $P_{B^{\prime}}$
to create a random sequence $P_{B^{\prime\prime}}=\Pi_{2Nn}P_{B^{\prime}}$
and sends that to Bob. The actual order is known to Alice only. 
\item [{DSQC4:}] After receiving Bob's authenticated acknowledgment of
receipt of all the qubits, Alice announces $\Pi_{Nn}\in\Pi_{2Nn}$,
the coordinates of the decoy qubits. The BB84 subroutine to detect
eavesdropping, is then implemented on the decoy qubits by measuring
them in the non-orthogonal bases $\left\{ |0\rangle,|1\rangle\right\} $
or $\left\{ |+\rangle,|-\rangle\right\} $. If sufficiently few errors
are found, then they go to the next step; else, they return to \textbf{DSQC1}.
\\
All intercept resend attacks will be detected in this step and
even if eavesdropping has happened Eve will not obtain any meaningful
information about the encoding operation executed by Alice as the
encoded sequence is rearranged. 
\item [{DSQC5:}] Alice discloses the coordinates of the remaining qubits. 
\item [{DSQC6:}] Bob measures his qubits in $\left\{ |b_{j}\rangle\right\} \equiv\left\{ |a_{\Pi(j)}\rangle\right\} $
basis and deterministically decodes the information encoded by Alice. 
\end{description}

\subsection{Turning the DSQC protocol to a QSDC protocol}

The above protocol is clearly a protocol of DSQC as Alice needs to
announce the actual order of the sequence. Rearrangement of particle
ordering present in the DSQC protocol may be avoided by sending the
information encoded states in $n$ steps and by checking eavesdropping
after each step. To be precise, consider that Alice sends a sequence
of all the first qubits first with $N$ decoy photons. If sufficiently
few errors are found, only then does she send the sequence of all
the second qubits, and so on. In such a situation the DSQC protocol
presented above will be reduced to a QSDC protocol as no classical
information will be required for decoding. Thus the previous protocol
can be easily generalized to a QSDC protocol. To do so we need to
modify \textbf{DSQC3}-\textbf{DQSC5} in the above protocol. Therefore,
the modified protocol may be described as follows: 
\begin{description}
\item [{QSDC1:}] Same as \textbf{DSQC1}. 
\item [{QSDC2:}] Same as \textbf{DSQC2}. 
\item [{QSDC3:}] Alice prepares $n$ sequences: $P_{Bs}=[p_{s},p_{s+n},\cdots,p_{s+(N-1)n}:1\leq s\leq n]$.
She also prepares $Nn$ decoy qubits as in \textbf{DSQC3} and inserts
$N$ of the decoy qubits randomly into each of the $n$ sequences
prepared by her. This creates $n$ extended sequences ($P_{B1+N},\, P_{B2+N},\cdots,\, P_{Bn+N}$)
each of which contains $2N$ qubits. Then she sends the first sequence
$P_{B1+N}$ to Bob. Receiving Bob's authenticated acknowledgment of
receipt of $2N$ qubits, she announces the positions of the decoy
qubits in $P_{B1+N}$. BB84 subroutine is then implemented on the
decoy qubits to check eavesdropping and if sufficiently few errors
are found then Alice sends $P_{B2+N}$ to Bob and they check for eavesdropping
and the process continues till the error free (i..e within the tolerance
limit) transmission of $P_{Bn+N}$. If at any stage of this step errors
more than the tolerable rate is detected then they truncate the protocol
and return to \textbf{QSDC1} else they go to the next step. 
\item [{QDDC4:}] Same as \textbf{DSQC6}. 
\end{description}
Since Eve cannot obtain more than 1 qubit of a $n$-partite state
(as we are sending the qubits one by one and checking for eavesdropping
after each step) she has no information about the encoded state and
consequently this direct quantum communication protocol is secure.
Thus the rearrangement of particle order is not required if we do
the communication in multiple steps. Further, this protocol does not
require any classical communication for the decoding operation. Consequently,
it is a QSDC protocol. Its qubit efficiency will be naturally higher
than the previous protocol. This is so because here Alice does not
need to disclose the actual sequence and consequently the amount of
classical communication required for decoding of the message is reduced.
But this increase in qubit efficiency is associated with a cost. This
QSDC protocol will be slower compared to its DSQC counterpart as Alice
has to communicate in multiple steps and has to check eavesdropping
in each of the steps.

\section{Generalization to orthogonal state based protocols\label{sec:Generalisation-to-orthogonal}}

The security of the above protocols arises from the non-commutativity
of the coding bases used in conjugate coding. However, GV protocol
employs orthogonal code states, and does not require conjugate coding.
In this section we will present a modified protocol that is independent
of conjugate coding. Or in other words we will represent a GV type
DSQC protocol using arbitrary states. In a completely orthogonal state
based protocol like GV protocol both error checking and encoding are
done by using orthogonal states. The DSQC and QSDC protocols presented
above already employ orthogonal encoding; however, error checking
is done by a BB84 subroutine, which uses non-orthogonal states (in
fact, MUBs). To turn the proposed DSQC and QSDC protocols into corresponding
GV-class protocols, which we call DSQC$^{{\rm GV}}$ and QSDC$^{{\rm GV}}$,
respectively, the error-checking in them also must employ only measurement
in a \textit{single} basis. In the following we will propose a new
trick, which will allow us to implement the error checking using the
Bell basis (alternatively by any other maximally entangled basis).

Intuitively, security in both the single-particle and multi-particle
case arises because of no-cloning and the fact that Eve cannot access
the full state at any given time. In Ref. \cite{With  preeti}, we
pointed out that whereas quantum duality \cite{englert} is the fundamental
origin of security in the GV protocol, monogamy of entanglement \cite{monogamy}
turns out to be that of security in the multi-particle GV protocols
obtained by us, suggesting, as a matter of considerable foundational
significance, that duality and monogamy are two sides so of the same
coin, a point to which we return later\textcolor{red}{{} }\cite{monastic}.

This unification is practically applied in the protocols presented
in this section, which arises fundamentally from the observation that
the security of the protocols we suggested above does not depend on
whether the basis states $|a_{j}\rangle$ are single-particle or multi-particle
states, as noted earlier. If $|a_{j}\rangle$ are taken as single-particle
states, we obtain a higher dimensional equivalent of the GV protocol.

\subsection{Turning DSQC to DSQC$^{{\rm GV}}$}

As the encoding in the DSQC protocol is already done with the orthogonal
states, so we retain the encoding steps (\textbf{DSQC1} and \textbf{DSQC2})
and the decoding steps (\textbf{DSQC5} and \textbf{DSQC6}) of the
\textbf{DSQC} and modify error checking steps \textbf{DSQC3} and \textbf{DSQC4}
as follows: 
\begin{description}
\item [{DSQC$^{{\rm GV}}$3:}] Using all the qubits of her possession,
Alice creates an ordered sequence $P_{B}=\left[p_{1},p_{2},p_{3},p_{4},\cdots,p_{Nn-1},p_{Nn}\right]$.
Now Alice prepares $\left[\frac{Nn}{2}\right]$ Bell pairs $|\psi^{+}\rangle^{\otimes\left[\frac{Nn}{2}\right]},$
where $|\psi^{+}\rangle=\frac{1}{\sqrt{2}}\left(|00\rangle+|11\rangle\right)$.
She uses the $Nn$ qubits of $|\psi^{+}\rangle^{\otimes\left[\frac{Nn}{2}\right]}$
as decoy qubits and adds them with $P_{B}$ to yield a larger sequence
$P_{B^{\prime}}=\left[p_{1},p_{2},p_{3},p_{4},\cdots,p_{Nn-1},p_{Nn},d_{1},d_{2},d_{3},d_{4},\cdots,d_{Nn-1},d_{Nn}\right]$.
Thereafter Alice applies a permutation operator $\Pi_{2Nn}$ on $P_{B^{\prime}}$
to create a random sequence $P_{B^{\prime\prime}}=\Pi_{2Nn}P_{B^{\prime}}$
and sends that to Bob. The actual order is known to Alice only. 
\item [{DSQC$^{{\rm GV}}$4:}] After receiving Bob's authenticated acknowledgment
of receipt of all the qubits, Alice announces $\Pi_{Nn}\in\Pi_{2Nn}$,
the coordinates of the decoy qubits (Bell states). Bob measures the
decoy qubits using Bell basis. If sufficiently few errors are found,
then they go to the next step; else, they return to \textbf{DSQC1}. 
\end{description}

\subsection{Turning QSDC to QSDC$^{{\rm GV}}$}

To turn the \textbf{QSDC} into \textbf{QSDC}$^{{\rm GV}}$ we just
need to modify \textbf{QSDC3} as follows (all the other steps will
remain same): 
\begin{description}
\item [{QSDC$^{{\rm GV}}$3:}] Alice prepares $n$ sequences: $P_{Bs}=[p_{s},\, p_{s+n},\cdots,p_{s+(N-1)n}:1\leq s\leq n]$.
She also prepares $\left[\frac{Nn}{2}\right]$ Bell pairs as in \textbf{DSQC$^{{\rm GV}}$3}
and inserts the qubits of the $\frac{N}{2}$ Bell pairs randomly into
each of the $n$ sequences prepared by her. This creates $n$ extended
sequences ($P_{B1+N},\, P_{B2+N},\cdots,\, P_{Bn+N}$) each of which
contains $2N$ qubits. Then she sends the first sequence $P_{B1+N}$
to Bob. Receiving Bob's authenticated acknowledgment of receipt of
$2N$ qubits, she announces the positions of the decoy qubits in $P_{B1+N}$.
Bob measures the decoy qubits using Bell basis to check eavesdropping
and if sufficiently few errors are found then Alice sends $P_{B2+N}$
to Bob and they check for eavesdropping and the process continues
till the error free (i.e. within the tolerance limit) transmission
of $P_{Bn+N}$. If at any stage of this step errors more than tolerable
rate is detected then they truncate the protocol and return to \textbf{QSDC$^{{\rm GV}}$1}
else they go to the next step. 
\end{description}
An interesting feature of these modified GV-type protocol is that
the strict time checking required for the original GV protocol is
not required here. This amplifies the experimental realizability of
the proposed orthogonal state based protocols. Further, in the original
GV protocol strictly one basis set is used for encoding, decoding
and eavesdropping checking. Same is true for modified version of GV
if $\{|a_{i}\rangle\}$ is Bell basis (or more precisely, if decoy
sates are prepared in $\{|a_{i}\rangle\}$ basis). In all other cases
encoding and decoding is done with a set of mutually orthogonal states
and eavesdropping is checked with another set of orthogonal states.
Extending this notion if we choose $\{|b_{j}\rangle\neq|a_{\Pi(j)}\rangle\}$
as an arbitrary output basis set then in the last step of the protocol
Bob has to do measurement in this output basis and we would obtain
GV-type protocol that uses three different sets of orthogonal basis
sets. Thus the presented protocols may use two/three different basis
sets to implement the protocol but they never use non-commutativity
of those two/three basis sets to obtain the unconditional security
of the protocol. This is so because eavesdropping is always checked
with a single basis as required in GV type protocol. Origin of security
in these protocols are explained below. In fact quantum mechanical
origin of the security of these protocols distinguishes them from
BB84 type of protocols.

\section{Efficiency and security of the protocol \label{sec:Efficiency-and-security}}

There are two aspects of the security of the DSQC and QSDC protocols:
1) We need to detect every possible attacks of Eve and 2) we need
to detect Eve before she obtains any meaningful information. The first
aspect is common in all secure quantum communication protocols (e.g.
QKD). Now, since we are inserting adequate number of decoy qubits
which are either prepared in a random sequence of $\{|0\rangle,|1\rangle,|+\rangle,|-\rangle\}$
in DSQC and QSDC schemes presented above, the eavesdropping checking
process is equivalent to BB84 protocol and consequently we can detect
all attacks of Eve. Similarly, when we inserts decoy qubits prepared
in Bell basis then the eavesdropping checking process is equivalent
to that in generalised GV protocol \cite{With preeti}. This step
is sufficient for QKD because if the key is leaked to Eve then the
key will not be used for encryption of any information in future but
this is not sufficient for DSQC or QSDC because here we are directly
sending information through the channel and we cannot afford to detect
Eve after she has already got all the information.

As the communicated states are orthogonal states in principle Eve
can intercept and resend them without being detected but the presence
of decoy photon will detect Eve's attacks. On the other hand, if Eve
attacks all qubits and measures them she will only obtain a random
sequence of bits. PoP ensures that she will not obtain any meaningful
information. Thus our protocol of DSQC with arbitrary states is unconditionally
secure.

An important point here is that security is obtained in general by
means of information splitting. First when we detect Eve, the entire
state may be publicly available, but the positions of decoy photons
are not announced till Bob receives all the photons. Without this
information it is impossible for Eve to get the string on which information
is encoded without being detected. But Eve would not mind being detected
after she obtains the information by an intercept-resend attack (substituting
dummies in place of real particles), because in that case she would
have already obtained the encoded message.

Here PoP plays a crucial role, making the encoded information to be
sent in essentially two or more pieces to Bob. Simultaneous non-availability
of these information pieces makes it impossible for Eve to get any
information. Thus the geographic information splitting plays an important
role in the security of our protocol.

More precisely, for a given state\textcolor{red}{{} }$|b_{j}\rangle$,
let the $M$ split pieces be denoted $|\alpha\rangle$, with $\alpha=1,2,\cdots,M$.
Assume that $|b_{j}\rangle\in\mathcal{H}_{d}$, the $d$-dimensional
Hilbert space of a single particle. Because of information splitting,
the most general interaction between Alice's transmitted system and
some probe with Eve is given by: \begin{equation}
\sum_{\alpha}|\alpha\rangle\langle\alpha|\otimes C_{\alpha},\label{eq:duality}\end{equation}
 where $C_{\alpha}$ are operations acting only on the probe, prepared
in an initial state $|0_{E}\rangle$. Intuitively, the more mutually
distinguishable states $|\beta_{j}\rangle\equiv C_{j}|0_{E}\rangle$,
the greater the entanglement generated between Alice's system and
the probe, and thus more mixed is the system received by Bob. This
is a manifestation of duality, which asserts a trade-off between the
`which-way' information Eve is effectively trying to get and the coherence
observed by Bob \cite{With preeti}. This disturbance, caused by Eve's
restricted power in interaction with the communication system, forms
the basis of security in GV. A rigorous derivation of duality in the
qubit case, as applied to the security of GV with qubits, is given
by us in Ref. \cite{With preeti} to be: \begin{equation}
\mathcal{P}+\mathcal{C}\le1,\label{eq:duality0}\end{equation}
 where $\mathcal{P}$ and $\mathcal{C}$ denote which-way information
and coherence, respectively.

In the multipartite case, a similar result holds, when we interpret
$|\alpha\rangle$ in Eq. (\ref{eq:duality}) to range over basis states
of individual particles. The resulting entanglement with the probe
causes a loss of coherence, which can be interpreted along the lines
of Eq. (\ref{eq:duality0}). By virtue of the convexity of entanglement,
this implies that the maximum entanglement or quantum correlation
that the system can form with the rest of the universe is lesser than
that which it could form if Eve's attack did not exist. More conventionally,
as we clarify elsewhere \cite{monastic}, this can be expressed as
the monogamy of entanglement \begin{equation}
E(A:C)+E(B:C)\le E(AB:C),\label{eq:monogamy}\end{equation}
 where $E(X:Y)$ is the square of concurrence, a measure of mixed-state
entanglement \cite{monogamy}.

The security of the extended GV-class protocols thus comes from duality
or monogamy, which are understood to be facets of the same underlying
phenomenon, which we have called \textit{monasticism} \cite{With preeti,monastic}.
One reflection of this is that security checking requires the use
of only a \textit{single} basis, unlike the case of BB84-class protocols
or the original QSDC or DSQC protocols, which require two or more
mutually unbiased bases. In the original GV protocol, this basis is
the $|0\rangle\pm|1\rangle$ basis, which can be generalized to an
appropriate `superposition' basis in the single-particle $d$-dimensional
generalization of GV. In the bipartite generalization of GV (both
in the QSDC or DSQC cases), this basis is typically that of the Bell
states. 

In the existing literature, two analogous but different parameters
are used for analysis of efficiency of DSQC and QSDC protocols. The
first one is simply defined as \begin{equation}
\eta_{1}=\frac{c}{q}\label{eq:efficency  1}\end{equation}
 where $c$ denotes the total number of transmitted classical bits
(message bits) and $q$ denotes the total number of qubits used \cite{Hwang-Hwang-Tsai,the:C.-W.-Tsai}.
This simple measure does not include the classical communication that
is required for decoding of information in a DSQC protocol. Consequently
it is a weak measure.

Another measure \cite{defn  of qubit  efficiency} that is frequently
used and which includes the classical communication is given as \begin{equation}
\eta_{2}=\frac{c}{q+b},\label{eq:efficiency
    2}\end{equation}
 where $b$ is the number of classical bits exchanged for decoding
of the message (classical communication used for checking of eavesdropping
is not counted). It is straightforward to visualize that $\eta_{1}=\eta_{2}$
for all QSDC and QSDC$^{{\rm GV}}$ protocols but $\eta_{1}>\eta_{2}$
for all DSQC protocols.

Now in the proposed \textbf{DSQC} and \textbf{DSQC}$^{{\rm GV}}$
protocols, $n$ bits of classical information are sent by $n-$qubits
and equal number (i.e. $n$) of decoy qubits so we have $c=n$ and
$q=2n$. Further to disclose the actual order we need $n$ bits of
classical information. Thus $b=n.$ Therefore, for DSQC and DSQC$^{{\rm GV}}$
protocols we have $\eta_{1}=\frac{1}{2}$ and $\eta_{2}=\frac{1}{3}$,
and similarly for \textbf{QSDC} and \textbf{QSDC$^{{\rm GV}}$} protocol
we have $\eta_{1}=\eta_{2}=\frac{1}{2}.$

These are the upper bounds on the qubit efficiency of this kind of
protocols, as shown in \cite{With Anindita-pla,With chitra-ijtp}.
As the generalized protocols proposed here uses arbitrary quantum
states to achieve these upper bounds, consequently, we may conclude
that the efficiency of DSQC and QSDC protocols of present kind is
independent of the choice of state. We have already shown that the
security of the protocol is also independent of the choice of state.
Consequently, all set of mutually orthogonal states are equivalent
as far as secure quantum communication is concerned. There is no advantage
of one state over the others and no essential advantage arises through
new DSQC and QSDC protocols that use complex states like 5-qubit Brown
state or 6-qubit cluster state.

\section{Conclusions\label{sec:Conclusions}}

It is shown that the DSQC and QSDC protocols of maximum efficiency
can be constructed by using any arbitrary linearly independent set
of quantum states applicable to single-particle or multi-particle
cryptography. Generalized protocols for the same are proposed and
their efficiency and security are analyzed. It is observed that the
proposed protocols are unconditionally secure and maximally efficient
for all states. Further, the protocols are generalized to completely
orthogonal state based protocols. This interesting idea is expected
to be of much use in all future experimental developments as it provides
a wide choice of states to experimentalists. Whereas the security
of conventional QSDC and DSQC protocols, like that of BB84-class QKD
protocols, is based on quantum non-commutativity, the security of
the GV versions of these protocols is based fundamentally on duality
(in the single-particle case) or monogamy of entanglement (in the
multi-particle case).

\textbf{Acknowledgment:} AP thanks Department of Science and Technology
(DST), India for support provided through the DST project No. SR/S2/LOP-0012/2010
and he also thanks the Operational Program Education for Competitiveness
- European Social Fund project CZ.1.07/2.3.00/20.0017 of the Ministry
of Education, Youth and Sports of the Czech Republic.

\end{document}